
\font\lbf=cmbx10 scaled\magstep2

\def\bs{\bigskip}
\def\ms{\medskip}

\def\ni{\noindent}
\def\cl{\centerline}

\def\title#1{\cl{\lbf #1}\ms}
\def\stitle#1{\bs{\ni\bf #1}\par\nobreak\ms}

\def\ref#1#2#3#4{#1\ {\it#2\ }{\bf#3\ }#4\par}
\def\refb#1#2#3{#1\ {\it#2\ }#3\par}
\def\CQG{Class.\ Qu.\ Grav.}
\def\GRG{Gen.\ Rel.\ Grav.}
\def\PR{Phys.\ Rev.}
\def\PRS{Proc.\ R.\ Soc.\ Lond.}

\def\paren#1{\left(#1\right)}
\def\abs#1{\left\vert#1\right\vert}
\def\o#1{\left.#1\right\vert_{t=0}}
\def\implies{\quad\Rightarrow\quad}
\def\nimplies{\quad\not\Rightarrow\quad}
\def\iff{\quad\Leftrightarrow\quad}
\def\D{{\cal D}}

\def\d{\delta}
\def\e{\varepsilon}
\def\t{\theta}
\def\s{\sigma}
\def\G{\Gamma}
\def\N{\nabla}
\def\p{\partial}
\def\half{{\textstyle{1\over2}}}
\def\quart{{\textstyle{1\over4}}}
\def\sixth{{\textstyle{1\over6}}}
\def\ud{{\dot u}}

\magnification=\magstep1

\title{Junction conditions for signature change}
\ms\cl{\bf Sean A. Hayward}
\ms\cl{Max-Planck-Institut f\"ur Astrophysik}
\cl{Karl-Schwarzschild-Stra\ss e 1}
\cl{85740 Garching bei M\"unchen}
\cl{Germany}
\ms\cl{Current address:}
\cl{Department of Physics}
\cl{Kyoto University}
\cl{Kyoto 606-01}
\cl{Japan}
\bs\cl{Revised 19th November 1994}
\bs\ni
{\bf Abstract.}
The change of signature of a metric
is explained using simple examples and methods.
The Klein-Gordon field on a signature-changing background is discussed,
and it is shown how the approach of Dray et al.\ can be corrected
to ensure that the Klein-Gordon equation holds.
Isotropic cosmologies are discussed,
and it is shown how the approach of Ellis et al.\ can be corrected
to ensure that the Einstein-Klein-Gordon equations hold.
A straightforward calculation shows that
a well defined Ricci tensor requires the standard junction condition,
namely vanishing of the second fundamental form of the junction surface.

\stitle{I. Introduction}\ni
The possibility that a metric can change its signature is an intriguing idea
which provides a classical model of quantum tunnelling in quantum cosmology.
With a degenerate metric,
it is not immediately clear that particular field equations admit solutions,
or even make sense.
Of course, familiar equations such as the Einstein equations do make sense
away from the junction between regions of different signature,
so that the problem reduces to
finding junction conditions for the relevant fields,
namely conditions which ensure that
the field equations make sense and are satisfied at the junction.
Unfortunately, several authors have attempted to describe signature change
without addressing the fundamental question of junction conditions,
but instead applying matching procedures
which happen to work for non-degenerate metrics.
This has led to various supposed ``solutions''
which do not actually satisfy the corresponding field equations.
The number of such papers is growing alarmingly.
The purpose of this article is to explain the topic of signature change
using simple examples and methods,
concentrating in particular on the points which have been incorrectly treated.
The Klein-Gordon field on a fixed background is treated in \S{}II
and homogeneous isotropic cosmologies in \S{}III,
using approaches that are as close as possible to those of
Dray et al.\ (1991, 1992, 1993) and Ellis et al.\ (1992) respectively.
In \S{}IV, a straightforward coordinate calculation is used to show that
a well defined Ricci tensor requires the standard junction conditions.

It is important to emphasize from the outset that
the junction conditions for a given field
(e.g.\ the Einstein gravitational field)
are not a question of taste,
nor determined by some ad hoc matching procedure,
but rather are determined by the relevant field equations
(e.g.\ the Einstein equations).
The field equations should make sense, and be satisfied, at the junction.
Making sense of the field equations can be rephrased
in a purely geometrical way as ensuring that
the corresponding geometrical objects (e.g.\ the Ricci tensor)
are well defined.

\stitle{II. The Klein-Gordon field}\ni
The simplest possible example is discussed first.
Consider a real Klein-Gordon field $\phi$ with zero potential,
on a 1-dimensional background with line-element $t\,dt^2$.
The field equation may be expressed in the covariant form
$$0=g^{ab}\N\!_a\N\!_b\phi=g^{ab}\paren{\p_a\p_b\phi-\G^c_{ab}\p_c\phi}
=t^{-1}\paren{\phi''-\half t^{-1}\phi'}\eqno(1)$$
or in partial derivative form either as
$$0=(\det g_{cd})^{-1/2}\p_a\paren{(\det g_{cd})^{1/2}g^{ab}\p_b\phi}
=t^{-1/2}\paren{t^{-1/2}\phi'}'\eqno(2)$$
or as
$$0=\abs{\det g_{cd}}^{-1/2}\p_a\paren{\abs{\det g_{cd}}^{1/2}g^{ab}\p_b\phi}
=\abs{t}^{-1/2}\paren{\e\abs{t}^{-1/2}\phi'}'\eqno(3)$$
or as
$$0
=\e\abs{\det g_{cd}}^{-1/2}\p_a\paren{\e\abs{\det g_{cd}}^{1/2}g^{ab}\p_b\phi}
=\e\abs{t}^{-1/2}\paren{\abs{t}^{-1/2}\phi'}'\eqno(4)$$
where $\phi$ is a function of $t$, the prime denotes $\p/\p t$
and $\e$ is the sign of $\det g_{cd}=t$.
In each case, the expressions are formal, due to the negative powers of $t$.
Making sense of equation (1) requires the asymptotic behaviour
$\phi'=O(t)$ and $\phi''-\half t^{-1}\phi'=O(t)$.
Similarly, making sense of equation (2), (3) or (4) requires
$\phi'=O(t^{1/2})$ and $(t^{-1/2}\phi')'=O(t^{1/2})$.
All four forms of the equation may then be multiplied out to give
$$0=t^{-1}\phi''-\half t^{-2}\phi'\eqno(5)$$
or equivalently
$$0=2t\phi''-\phi'.\eqno(6)$$
This is manifestly well defined if $\phi$ is twice differentiable.
All these forms of the equation imply
$$0=\paren{t^{-1}\phi''-\half t^{-2}\phi'}2\phi'
=\paren{t^{-1}(\phi')^2}'\eqno(7)$$
which is the Klein-Gordon equation in conservation form,
$0=\N^aT_{ab}$.
This integrates to
$$(\phi')^2=ct\eqno(8)$$
for constant $c$. Since $t$ changes sign and $(\phi')^2\ge0$,
it follows that $c=0$, so that there is only the trivial solution
$$\phi'=0.\eqno(9)$$
This is a crucial point: normally the initial data for a second-order
ordinary differential equation are $\o{(\phi,\phi')}$.
For this {\it singular} equation, the initial data are just $\o\phi$,
with the restriction $\o{\phi'}=0$.
The equation is an archetype of the singular partial differential equations
that occur under signature change in higher dimensions,
where the usual initial data $\o{(q,q')}$ are restricted by $\o{q'}=0$.
It is these restrictions or junction conditions which
some authors claim are unnecessary.

Dray et al.\ (1991) consider the above problem, claiming that
the covariant and partial derivative forms of the equation
are essentially different,
and that the general solution is any one of
$$\phi'=At^{1/2}\qquad
\phi'=A\abs{t}^{1/2}\qquad
\phi'=A\e\abs{t}^{1/2}\eqno(10)$$
depending on the form of the equation and the method of solution.
They claim that the last ``solution'' is more fundamental.
The meaning of this is unclear.
The only justification given is that
the last ``solution'' admits a conserved inner product,
but this is not true of other spin fields (Romano 1993).

In general, the junction condition for the Klein-Gordon equation
under a change of signature,
apart from continuity of $\phi$,
is that the momentum $\psi$ conjugate to $\phi$
satisfies $\o\psi=0$.
This junction condition on signature-changing solutions,
referred to also as real tunnelling solutions,
is well known and regarded as standard in the context of quantum cosmology,
e.g.\ Halliwell \& Hartle (1990), Gibbons \& Hartle (1990),
Fujiwara et al.\ (1991),
Carlip (1993), Hawking et al.\ (1993), Vilenkin (1993),
Barvinsky \& Kamenshchik (1993).
The familiar argument is that
the momentum fields are real in the Lorentzian region
and imaginary in the Riemannian region,
and so must vanish at the junction.
Other derivations, using various different approaches,
have been given by Hayward (1992a) and Kossowski \& Kriele (1993ab, 1994abc).
One simple way to see this condition is as follows.
Take normal coordinates $(t,x^i)$ such that
$$ds^2=-t\,dt^2+h_{ij}dx^idx^j.\eqno(11)$$
On this background, the Klein-Gordon equation follows from Hayward (1992) as
$$\phi''-\half t^{-1}\phi'=-\t\phi'+t(\D^2\phi-V_\phi)\eqno(12)$$
where $\t=\half h^{ij}h'_{ij}$ is the expansion,
$\D^2$ is the Laplacian of $h_{ij}$,
and $V_\phi$ is the derivative of the potential $V(\phi)$.
Note that this is equation (1) with a source term.
This is a simple example of the singular partial differential equations
that occur in the context of signature change.
A general existence and uniqueness theorem for such systems
has recently been given by Kossowski \& Kriele (1994b).
For the moment, it suffices to note that the equation involves $\phi''$,
and so makes sense if $\phi$ is twice differentiable.
The singular $t^{-1}$ term in the Klein-Gordon equation
determines the asymptotic behaviour
$$\phi'=O(t).\eqno(13)$$
This implies that the momentum $\psi=t^{-1/2}\phi'$
is well defined and satisfies $\o\psi=0$,
as noted previously (Hayward 1992a).

Other papers based on the flawed analysis of Dray et al.\ (1991)
include Dray et al.\ (1992, 1993), Romano (1993) and Hellaby \& Dray (1994).
In particular, Dray et al.\ (1993) have claimed to justify
their previous assertions by deriving junction conditions of their own.
It turns out that these junction conditions are necessary but not sufficient
to ensure that the field equations are well defined,
as follows.
In the language of differential forms,
the Klein-Gordon equation can be written as
$$*d{*}d\phi=V_\phi.\eqno(14)$$
Dray et al.\ take the vacuum equation, or wave equation, in the form
$$d{*}d\phi=0\eqno(15)$$
and consider joining solutions across a spatial hypersurface $t=0$.
Denote the regions $t>0$ and $t<0$ by $U_+$ and $U_-$ respectively,
and $[F]=\lim_{t\to0_+}F-\lim_{t\to0_-}F$.
If the metric is non-degenerate,
a standard distributional method shows that
the necessary and sufficient junction conditions for the wave equation (15) are
$[\phi]=0$ and $[*d\phi]\wedge dt=0$, as follows.
Introduce the Heaviside distributions $\t_\pm$
with support on $U_\pm$,
and the Dirac distibution $\d=\pm d\t_\pm$.
Assume that the restrictions $\phi\vert_{U_\pm}$ are $C^\infty$,
and extend them to $C^\infty$ functions $\phi_\pm$
on $M=\overline{U}_+\cup\overline{U}_-$.
Then on $M$, $\Phi=\t_+\phi_++\t_-\phi_-$ is a distribution
in the sense of Schwartz (1950),
and coincides with $\phi$.
The wave equation (15) then coincides with
$$0=d{*}d\Phi=\t_+d{*}d\phi_++\t_-d{*}d\phi_-+2\d\wedge[*d\phi]+[\phi]d{*}\d.
\eqno(16)$$
Since the distributions $(\t_\pm,\d,d{*}\d)$ are linearly independent,
this is equivalent to
$$d{*}d\phi\vert_{U_\pm}=0\qquad
[*d\phi]\wedge dt=0\qquad
[\phi]=0.\eqno(17)$$
However, if there is a change of signature,
the problem is completely different,
because the operator $*$ is then {\it singular} when acting on 1-forms $d\phi$.
For instance, taking normal coordinates (11),
$*d\phi$ contains a factor of
$t^{-1/2}$, $\e\abs{t}^{-1/2}$ or $\abs{t}^{-1/2}$
multiplying $\phi'$,
as in equation (2), (3) or (4) respectively.
The standard method does not then make sense.
This is easily corrected:
the operator $*$ may be interpreted as
a distribution in the sense of Schwartz (1950),
since $\e\abs{t}^n$ can be interpreted as distributions:
$\e\abs{t}^n=t_+^n-t_-^n$
in terms of the Schwartz monomial distributions $t_\pm^n$
(Gel'fand \& Shilov 1964).
The wave equation (15) then makes sense
in the context of the Schwartz algebra if and only if $d\phi$ is $C^\infty$.
Since $\phi'=O(t)$ one finds the junction condition $\o\psi=0$,
where $\psi=\e\abs{t}^{-1/2}\phi'$.
Equivalently,
$$\o{*d\phi}\wedge dt=0.\eqno(18)$$
In contrast, Dray et al.\ (1993) apply the standard distributional method to
the equation $dF=0$,
obtaining the junction condition $[F]\wedge dt=0$.
They make no attempt to derive junction conditions for the equation $F=*d\phi$,
or even to make sense of it,
but simply state that $\phi$ should be continuous.
Thus they do not address the key problem of making sense of
the singular nature of the wave equation under a change of signature.
In summary,
Dray et al.\ have found {\it necessary} (and undisputed) junction conditions
for the wave equation (15),
whereas {\it sufficient} junction conditions would be required
to justify their claim\footnote\dag
{The abstract of Dray et al.\ (1993) states that:
``Reformulating the problem using junction conditions,
we then show that the solutions obtained above are the unique ones
which satisfy the natural distributional wave equation {\it everywhere}.''
Similarly, the abstract of Dray et al.\ (1991) states that:
``Choosing the wave equation so that
there will be a conserved Klein-Gordon product implicitly determines
the junction conditions one needs to impose in order to obtain
{\it global solutions}.'' (My italics).}
to have obtained solutions which satisfy the wave equation everywhere.

Hellaby \& Dray (1994) claim
that conservation laws fail at a change of signature.
This is manifestly untrue: a straightforward formal calculation
yields the contracted Bianchi identity $\N^aG_{ab}=0$.
The Einstein equation $G_{ab}=T_{ab}$
then implies the conservation equation $\N^aT_{ab}=0$.
Similarly, the Klein-Gordon equation $\N^2\phi=0$
independently implies the conservation equation:
$$\N^aT_{ab}=\N^a\half(\N_a\phi\N_b\phi-\half g_{ab}g^{cd}\N_c\phi\N_d\phi)
=\half(\N_b\phi)\N^2\phi=0.\eqno(19)$$
Hellaby \& Dray have thus inadvertently demonstrated the absurdity
of the claim that the junction conditions need not hold:
by violating the conservation equation they violate the field equations.
The junction conditions restore the field equations
and consequently the conservation equation (Hayward 1994b).

\stitle{III. Homogeneous isotropic cosmologies}\ni
The second simple example of signature change is provided by
homogeneous isotropic cosmologies,
for which the line-element may be written as
$$ds^2=-N\,dt^2+a^2d\Sigma^2\eqno(20)$$
where $d\Sigma^2$ refers to a constant-curvature space,
and the scale factor $a$ and squared lapse $N$ are functions of $t$.
For simplicity, take the matter model to be a scalar field $\phi$
with potential $V(\phi)$.
The Einstein-Klein-Gordon equations, with or without a change of signature,
have been given in a ``3+1'' decomposition previously (Hayward 1992a),
and for the line-element (20) reduce to
$$\eqalignno{&N\phi''-\half N'\phi'=-3Na^{-1}a'\phi'-N^2V_\phi&\cr
&Na''-\half N'a'=-\sixth Na(\phi')^2+\sixth N^2aV&\cr
&0=12(a')^2-(a\phi')^2+2N(6k-a^2V)&(21)\cr}$$
where the prime denotes $\p/\p t$,
and $k=-1$, 0, 1 labels the hyperbolic, flat and spherical cases respectively.
These are the Klein-Gordon, Raychaudhuri and Friedmann equations respectively,
valid whether the signature changes or not.
Ellis et al.\ (1992) give the same equations
except for an overall factor in $\phi$.
The convention here is that the kinetic energy is $\half(\phi')^2$
and the potential energy is $V$.

Consider a change of signature described by $N$ changing sign at $t=0$.
Ellis et al.\ choose a discontinuous lapse, $N=\e$,
where $\e$ is the sign of $t$.
Making sense of this discontinuous-metric approach
requires the use of distributions,
since the evolution equations involve $N'$.
Specifically,
one needs the Dirac distribution $\d$ with support on the junction surface,
since $\e'=2\d$.
Both $\e$ and $\d$ are distributions in the sense of Schwartz (1950).
Direct substitution of $N=\e$ into the field equations (21) yields
$$\eqalignno{&\e\phi''-\d\phi'=-3\e a^{-1}a'\phi'-\e^2V_\phi&\cr
&\e a''-\d a'=-\sixth\e a(\phi')^2+\sixth\e^2aV&\cr
&0=12(a')^2-(a\phi')^2+2\e(6k-a^2V).&(22)\cr}$$
Since the distributions $(\e,\d)$ are linearly independent,
the equations divide into the regular parts
$$\eqalignno{&\e\phi''=-3\e a^{-1}a'\phi'-\e^2V_\phi&\cr
&\e a''=-\sixth\e a(\phi')^2+\sixth\e^2aV&\cr
&0=12(a')^2-(a\phi')^2+2\e(6k-a^2V).&(23)\cr}$$
and the singular parts
$$\phi'\d=0\qquad a'\d=0.\eqno(24)$$
The latter are the junction conditions,
expressing the vanishing of $(\phi',a')$ at the junction.
This special case of the general result (Hayward 1992a)
has been emphasised explicitly (Hayward 1992b).
It is important to emphasize that in this approach,
the junction conditions are not {\it assumed}
in order to make sense of the field equations,
which are already well defined in a distributional sense,
but are {\it derived}
as the singular part of the field equations.

Ellis et al.\ (1992) take $N=\e$,
but when substituting this choice into the field equations (21),
obtain only the regular parts (23) whilst omitting the singular parts (24).
It is easily checked by substitution that their ``solutions''
do not satisfy the field equations,
except for a previously known case (Hayward 1992ab).
This has been confirmed in great detail by Kossowski \& Kriele (1993b).

The same basic error has been repeated by
Ellis (1992), Kerner \& Martin (1993),
Ellis \& Piotrkowska (1994), Carfora \& Ellis (1994) and Embacher (1994ab).
Unfortunately, Ellis et al.\ insist that
their ``solutions'' do satisfy the field equations.\footnote\ddag
{E.g.\ in the framed remark in the conclusion of Ellis et al.\ (1992):
``We can construct explicit solutions of the field equations
where there is a change of signature...
the solution is without surface layer terms.''
Or in \S5: ``[The continuity conditions]
{\it ensure the field equations continue through the transition}.''
(Their italics).
Or the opening remark of the conclusion of Ellis (1992):
``The Einstein equations, formulated in a 3+1 way,
allow a change of signature without divergences occurring
and without surface layers.''}
They give many arguments intended to support this claim,
the most lucid of which are examined below.
The paper of Ellis et al.\ (1992) will be considered;
many of the erroneous arguments are repeated in the later papers,
as explained briefly at the end of this section.

Initially, Ellis et al.\ (\S1--2) consider a general squared lapse $N$,
and remark on the difference between
the Lie derivative of functions $f$ along the vector $\p/\p t$,
$f'=\p f/\p t$,
and the affine derivative
$$\dot f=\abs{N}^{-1/2}f'.\eqno(25)$$
They accept that $(\phi',a')$ vanish at $t=0$,
but argue that $(\dot\phi,\dot a)$ need not,
due to the singular factor in $N$.
That is, they implicitly assert the asymptotic behaviour
$(\phi',a')=O(\abs{N}^{1/2})$ as $N\to0$.
However, the asymptotic behaviour is determined by the field equations,
as follows.
The evolution equations involve the derivatives $(\phi'',a'')$,
and so make sense if $(\phi,a)$ are twice differentiable.
The field equations (21) determine the asymptotic behaviour
$$(\phi',a')=O(N)\eqno(26)$$
as $N\to0$, if $N'$ is non-zero.
Thus the conditions
$$\o{(\dot\phi,\dot a)}=0\eqno(27)$$
are obtained.

Ellis et al.\ (\S2--3) also distinguish between the affine parameter $s$
defined by
$$ds=\sqrt{\e N}dt\eqno(28)$$
and another parameter $\s$ supposedly defined by
$d\s=\sqrt{N/\e}dt$.
They claim (\S3) that $\s$ is well defined while $s$ is not;
the opposite is manifestly true, since $1/\e$ is undefined.
This means that the dynamical variables which Ellis et al.\ use
are undefined at the junction, strictly speaking.
This is easily corrected by replacing $\s$ with $s$.
The use of $\s$ appears to be tied up with
attempts to argue the distributional terms away,
but these arguments are unclear.

Kossowski \& Kriele (1993a) have shown that
coordinates exist such that $N=t$.
Ellis et al.\ (\S3) argue that $N=t$ is not a good choice of coordinates,
and that a suitable choice for $N$ is $\e t^2$, or higher powers of $t$.
The meaning of this is unclear.

Ellis et al.\ (\S4--5) claim that $N=t$ and $N=\e$ represent
different differential structures.
Presuming that this means that
they correspond to two different manifolds on the same topological space,
this cannot be correct,
since there is only one such differential structure on the real line,
usually represented by a function with non-vanishing derivative.
In terms of the line-element (20),
$N=t$ is such a function and $N=\e$ is not.
Ellis et al.\ also claim that different continuous functions $N(t)$
represent different differential structures.
In fact, any differentiable function $N$ of $t$ with differentiable inverse
would represent the same differential structure as $N=t$.
A precise statement of equivalence between the $N=t$ and $N=\e$ approaches
has recently been given by Kossowski \& Kriele (1993b).

Having eventually settled on the choice $N=\e$ as being what physics suggests,
Ellis et al.\ (\S5) claim that
there are no distributional contributions to the curvature,
on the grounds that the jump $[\G]$ in the connection $\G$ vanishes.
However, $\p_0g_{00}=-2\d$,
which gives the formal expression $\G^0_{00}=-\d/\e$.
This is undefined, strictly speaking,
but clearly much worse than a jump discontinuity in $\G$.
The consequent distributional terms in the curvature tensor
appear explicitly in the field equations (22).
The mistake appears to be due to
a misquotation of the analysis of junction conditions by Barrab\`es (1989),
who demands a continuous metric.

It should also be remarked that having chosen $N=\e$,
the affine parameter is given by (28) as $s=t$, fixing the zero.
Since now $\dot f=f'$ for any function $f$,
and Ellis et al.\ do accept that $(\phi',a')$ vanish at the junction,
it is unclear why they believe that $(\dot\phi,\dot a)$ do not vanish.

Ellis et al.\ (\S5) argue that their ``solutions'' are justified
by appealing to the junction conditions of Darmois (1927)
as opposed to Lichnerowicz (1955).
The Darmois conditions impose continuity of
the first and second fundamental forms of the junction,
in this case $(a,\dot a)$,
while the Lichnerowicz conditions impose continuity of
the metric and its first derivatives,
in this case $(a,N,a',N')$.
There are also what will be called the Hamiltonian conditions
(e.g.\ Barrab\`es 1989, Clarke \& Dray 1987):
continuity of the metric (first fundamental form, lapse and shift)
and the second fundamental form,
in this case $(a,N,\dot a)$.
These are all intended as necessary junction conditions
which in vacuum are also sufficient,
though the Hamiltonian conditions are easily generalized
to fields defined by a Hamiltonian.
One should also distinguish between
these conditions in particular coordinates $x$,
denoted by D$(x)$, L$(x)$ and H$(x)$,
and the corresponding statements of existence,
$\exists x$: D$(x)$, L$(x)$ and H$(x)$,
denoted by D, L and H, respectively.
For the particular forms,
there is a strict hierarchy if the metric is non-degenerate:
$$\hbox{L}(x)\implies\hbox{H}(x)\implies\hbox{D}(x)
\nimplies\hbox{H}(x)\nimplies\hbox{L}(x).\eqno(29)$$
Of these, H$(x)$ is optimal in that it imposes continuity of
the configuration and momentum variables in the Hamiltonian formulation,
which are the quantities that actually occur in
a first-order form of the vacuum Einstein equations
(e.g.\ Fischer \& Marsden 1979).
For the existential forms of the conditions,
$$\hbox{L}\iff\hbox{H}\iff\hbox{D}\eqno(30)$$
if the metric is non-degenerate, since then there exist normal coordinates,
i.e.\ with unit lapse and zero shift (Bonnor \& Vickers 1981).
So in the context for which they were intended, L, H and D are equivalent.
However, normal coordinates do not exist across a change of signature,
where the lapse is zero,
and a standard coordinate system is instead given by
zero shift and $N=t$ (Kossowski \& Kriele 1993a).
Ellis et al (\S5) emphasize that their discontinuous-metric approach
satisfies D$(x)$ but not L$(x)$ for their coordinates $x$,
and conclude that this justifies non-zero $(\dot\phi,\dot a)$ at the junction.
Nevertheless,
the junction conditions follow as the singular part of the field equations,
as explained above.
In other words, even if one assumes only the Darmois conditions
(in the form D$(x)$ or D),
the field equations imply the full junction conditions (24).
The moral is that one cannot
simply quote junction conditions formulated for non-degenerate metrics
and apply them uncritically to degenerate metrics;
in either case,
the correct junction conditions are determined by the field equations.

Similar mistakes to those of Ellis et al.\ (1992) occur in more recent papers.
In particular, Ellis (1992) constructs a formalism for signature change
involving an evolution vector $u=\p/\p t$ with discontinuous normalization,
$u_au^a=-\e$.
When deriving the field equations, Ellis takes the choice $\ud^a=0$,
and implicitly also assumes $\ud_a=0$.
This is inconsistent with the normalization,
which differentiates to $\ud_au^a+u_a\ud^a=-2\d$.
In this way, the singular parts of the field equations are again missed.

An almost identical mistake occurs in the papers of Embacher (1994ab),
who also reports signature-changing ``solutions''
which do not actually satisfy the relevant field equations.
By adopting an approach in which the field equations are ill defined
at the junction,
Embacher obtains equations analogous to (23) rather than (22),
i.e.\ missing the terms in $\delta$.
Embacher also claims
that Hayward (1992a) disagrees with Ellis et al.\ (1992)
because they adopt different approaches,
whereas the former reference actually considers
the discontinuous-metric approach of Ellis et al.,
deriving the junction conditions that Ellis et al.\ missed.

Carfora \& Ellis (1994) claim to justify the previous claims of Ellis et al.,
in particular repeating the claim that ``no surface layer is present''
in their ``solutions''.
The mistake of Carfora \& Ellis is that
they consider only the constraint equations and ignore the evolution equations.
The distributional (or surface layer) terms occur in the evolution equations,
as shown above and in previous papers
(Hayward 1992a, Kossowski \& Kriele 1993b).
Carfora \& Ellis also claim
that the latter references disagree with Ellis et al.\
because they adopt different approaches,
whereas both references actually consider
the discontinuous-metric approach of Ellis et al.,
deriving the junction conditions that Ellis et al.\ missed.

Paradoxically, Carfora \& Ellis admit at one point
that the field equations do require the junction conditions,
without citing the appropriate references and without mentioning
that this contradicts the claims of Ellis et al.\ (1992).
Instead, they conclude that this justifies the claims of Ellis et al.
They argue that it is better not to satisfy the field equations.
This new philosophy is not mentioned
in the abstract, introduction or conclusion,
which instead contain repeated claims that the junction conditions arise from
``continuity suppositions'', ``smoothness assumptions'',
a ``more restricted approach'', ``more stringent restrictions'',
``additional differentiability'' or ``extra conditions'',
rather than just the field equations.

\stitle{IV. The Riemann tensor}\ni
Kossowski \& Kriele (1994abc)
have shown by purely geometrical methods that
the Riemann tensor can be continued smoothly across a change of signature
if and only if two conditions are met,
one of which is the standard junction condition
that the second fundamental form vanish,
and the other of which is the limit of the evolution equations at the junction.
In this section,
these results are derived by a straightforward coordinate method.

Kossowski \& Kriele (1993a) show that if the metric changes signature
from Riemannian to Lorentzian across a spatial surface,
then it is possible to take normal coordinates $(x^0,x^i)$
such that the metric is
$$g_{00}=-t\qquad g_{0i}=0\qquad g_{ij}=h_{ij}\eqno(31)$$ with $t=x^0$.
Denote $f'=\p_0f$.
The connection divides into the regular components
$$\G^i_{00}=0\qquad\G^0_{0i}=0\qquad
\G^i_{0j}=\half h^{ik}h'_{jk}=O(1)\qquad\G^i_{jk}=O(1)\eqno(32)$$
and the singular components
$$\G^0_{ij}=\half t^{-1}h'_{ij}\qquad\G^0_{00}=\half t^{-1}.\eqno(33)$$
Thus the covariant derivative is singular.
Nevertheless, the Riemann tensor is regular under suitable junction conditions.
(A regular tensor is one which is actually a tensor,
according to the usual definition).
The Ricci tensor takes the form
$$\eqalignno{R_{00}&=O(1)&\cr
R_{0i}&=O(1)&\cr
R_{ij}&=\p_0\G^0_{ij}+\G^0_{00}\G^0_{ij}-\G^0_{ki}\G^k_{j0}-\G^0_{kj}\G^k_{i0}
+O(1)&\cr
&=-\quart t^{-2}h'_{ij}+\half t^{-1}(h''_{ij}-h^{km}h'_{ik}h'_{jm})
+O(1).&(34)\cr}$$
Demanding a regular Ricci tensor therefore yields the conditions
$$h'_{ij}=O(t)\qquad
h''_{ij}-\half t^{-1}h'_{ij}=O(t).\eqno(35)$$
It follows that
$$\G^i_{0j}=O(t)\qquad\G^0_{ij}=O(1).\eqno(36)$$
With these conditions, the Weyl tensor is also regular,
$$C_{\mu\nu\rho}{}^\sigma=O(1).\eqno(37)$$
The reason is that the singular connection component $\G^0_{00}$
does not enter the Riemann tensor as $\p_0\G^0_{00}$ or $(\G^0_{00})^2$,
due to symmetries,
but only as above in $R_{ij}$, or multiplied by something $O(t)$.
This does not provide a very useful formalism,
since in practice one wishes to know exactly what the $O(1)$ terms are,
and these appear as long unilluminating expressions in components.
Writing these expressions in a geometrically clear way
is precisely the purpose of the ``3+1'' formalism (Hayward 1992a).
However, this does provide a straightforward,
purely geometrical derivation of the junction conditions.

Note that the conditions (35) imply
$$\o{(\e\abs{t}^{-1/2}h'_{ij})}=0\qquad
\o{(\e\abs{t}^{-1/2}h'_{ij})'}=0.\eqno(38)$$
Equivalently, $$\o{p^*}=0\qquad\o{(p^*)'}=0\eqno(39)$$
or $$\o{P}=0\qquad\o{P'}=0\eqno(40)$$
where $p^*$ is as in \S2 and $P$ as in \S5 of Hayward (1992a).
In each case, the first equation is the junction restriction,
i.e.\ the constraint on the initial data at the junction,
with the second equation being the limit of the evolution equations.
In other words,
the first condition is necessary for the Einstein equations
to be well defined,
and the second condition is necessary for the Einstein equations
to be satisfied.
Kossowski \& Kriele (1994abc) give rigourous proofs of these conditions,
which correspond to the vanishing of their tensors
$I\!\!I$ and $I\!\!I\!\!I$ respectively.

\stitle{V. Remarks}\ni
In the context of quantum cosmology,
the junction conditions on signature-changing solutions,
or real tunnelling solutions,
are usually regarded as obvious:
the momentum fields are real in the Lorentzian region
and imaginary in the Riemannian region,
and so must vanish at the junction (e.g.\ Gibbons \& Hartle 1990).
Unfortunately, the commonly used Wick-rotation argument ($t\mapsto it$)
is simply invalid as a justification for
joining two spaces of different signature.
To describe the change of signature in a geometrically respectable way,
one requires real coordinate charts covering the junction.
Normally the fields are required to be ordinary tensor functions
on the underlying manifold,
though distributions can also be allowed, as in \S{}III.
The fields themselves may still be complex,
though the geometrical meaning of an imaginary second fundamental form
is likely to be questioned.
Various ways to avoid this were described previously (Hayward 1992a):
take moduli of the momenta,
and take care with the consequent distributional terms;
work with the squared momenta;
or work directly with the second-order form of the field equations.
The last approach has the merit that the problem becomes one of
finding suitably smooth solutions to singular differential equations,
and has been significantly developed by Kossowski \& Kriele (1993ab, 1994abc).
Dereli \& Tucker (1993) have also suggested such an approach,
but did not specifically note the consequent junction conditions.
It should come as no surprise to find that the same junction conditions
are found in all these approaches, though apparently for different reasons,
according to whether one is thinking in terms of
field theory, geometry, differential equations or distributions.

It should also be remarked that the signature-changing solutions
are rather special from the viewpoint of quantum cosmology,
where they correspond to solutions which, in complex momentum space,
run along the imaginary axis and then switch to the real axis at the origin.
This already involves a lack of smoothness which one might prefer to avoid,
for instance by considering smooth curves which lie close to the axes.
All that is really required of a tunnelling solution is that
it be a leading-order (in $\hbar$) solution to the Wheeler-DeWitt equation,
asymptotes to a Lorentzian solution for consistency with observation,
and is compatible with one's preferred boundary condition
(e.g.\ Hawking et al.\ 1993, Vilenkin 1993, Barvinsky \& Kamenshchik 1993).
The sort of signature-changing ``solutions''
that Dray et al.\ and Ellis et al.\ propose correspond to curves
that jump discontinuously from the imaginary axis to the real axis,
which is incompatible with the dynamics.
So from the viewpoint of quantum cosmology,
even if such ``solutions'' admit a modified equation
to which they are a solution,
this would still be regarded as a mathematical curiosity.
This remark is relevant because it turns out to be possible to find
weak solutions across a change of signature
which are not solutions in the usual sense (Hayward 1994a).

The restrictiveness of the junction conditions means that
signature change is not a classically stable phenomenon.
This is just as well, since otherwise
one would have to explain why it is not continually observed.
It seems to be this restrictiveness
which has led to the mistaken attempts to avoid the conditions.
Actually,
the restrictiveness can be taken as a strength in that it yields predictions:
if something like a change of signature did occur around the Planck epoch,
as in the quantum-tunnelling scenario for the birth of the universe,
then the possible universes are severely restricted
as compared to the standard initial-singularity scenario.
Remarkably, these restrictions seem to agree well with what is required
to explain the special features of the observed universe (Hayward 1993).
\bs\ni
Acknowledgements.
It is a pleasure to thank Marcus Kriele for discussions.
Sections II and III arose due to correspondence with
Tevian Dray and George Ellis respectively.
I am grateful to the Max-Planck-Gesellschaft
and the Japan Society for the Promotion of Science for financial support.
\bs
\begingroup
\parindent=0pt\everypar={\global\hangindent=20pt\hangafter=1}\par
{\bf References}\par
\ref{Barrab\`es C 1989}\CQG6{581}
\refb{Barvinsky A O \& Kamenshchik A Y 1993}{Tunnelling geometries I}
{(gr-qc/9311022)}
\ref{Bonnor W B \& Vickers J A 1981}\GRG{13}{29}
\refb{Carfora M \& Ellis G 1994}
{The geometry of classical change of signature}{(gr-qc/9406043)}
\ref{Carlip S 1993}\CQG{10}{1057}
\ref{Clarke C J S \& Dray T 1987}\CQG4{265}
\refb{Darmois G 1927}{M\'emorial des Sciences Math\'ematiques}
{Fasc.~25 (Paris: Gauthier-Villars)}
\ref{Dereli T \& Tucker R W 1993}\CQG{10}{365}
\ref{Dray T, Manogue C A \& Tucker R W 1991}\GRG{23}{967}
\refb{Dray T, Manogue C A \& Tucker R W 1992}
{The effect of signature change on scalar field propagation}
{(Oregon preprint)}
\ref{Dray T, Manogue C A \& Tucker R W 1993}\PR{D48}{2587}
\ref{Ellis G F R 1992}\GRG{24}{1047}
\refb{Ellis G F R \& Piotrkowska K 1994}
{Proc.\ Les Journ\'ees Relativistes}{(to appear)}
\ref{Ellis G, Sumeruk A, Coule D \& Hellaby C 1992}\CQG{9}{1535}
\refb{Embacher F 1994a}
{Signature change induces compactification}{(gr-qc/9410012)}
\refb{Embacher F 1994b}
{The trace left by signature-change-induced compactification}
{(gr-qc/\-9411028)}
\refb{Fischer A E \& Marsden J E 1979 in}
{General Relativity, an Einstein Centenary Survey}
{ed: Hawking S W \& Israel W (Cambridge University Press)}
\ref{Fujiwara Y, Higuchi S, Hosoya A, Mishima T \& Siino M 1991}\PR{D44}{1756}
\refb{Gel'fand I M \& Shilov G E 1964}{Generalized Functions}
{(New York: Academic Press)}
\ref{Gibbons G W \& Hartle J B 1990}\PR{D42}{2458}
\ref{Halliwell J J \& Hartle J B 1990}\PR{D41}{1815}
\ref{Hawking S W, Laflamme R \& Lyons G W 1993}\PR{D47}{5342}
\ref{Hayward S A 1992a}\CQG9{1851; erratum 2453}
\refb{Hayward S A 1992b}
{Comment on ``Change of signature in classical relativity''}{(unpublished)}
\ref{Hayward S A 1993}\CQG{10}{L7}
\ref{Hayward S A 1994a}\CQG{11}{L87}
\refb{Hayward S A 1994b}{Comment on ``Failure of standard conservation laws
at a classical change of signature''}{(preprint)}
\ref{Hellaby C \& Dray T 1994}\PR{D49}{5096}
\ref{Kerner R \& Martin J 1993}\CQG{10}{2111}
\ref{Kossowski M \& Kriele M 1993a}\CQG{10}{1157}
\ref{Kossowski M \& Kriele M 1993b}\CQG{10}{2363}
\ref{Kossowski M \& Kriele M 1994a}\PRS{A444}{297}
\ref{Kossowski M \& Kriele M 1994b}
{The Einstein equation for signature-type-changing space-times, \PRS}A
{(to appear)}
\refb{Kossowski M \& Kriele M 1994c}{Characteristic classes for
transverse type-changing pseudo-Riemannian metrics with smooth curvature}
{(preprint)}
\refb{Lichnerowicz A 1955}
{Th\'eories Relativistes de la Gravitation et de l'\'Electromagn\'etisme}
{(Paris: Masson)}
\ref{Romano J D 1993}\PR{D47}{4328}
\refb{Schwartz L 1950}{Th\'eorie des Distributions}{(Paris: Hermann)}
\refb{Vilenkin A 1993}{Quantum cosmology}{(gr-qc/9302016)}
\endgroup
\bye